\documentclass{IEEEtran}
\usepackage{cite}
\usepackage{xcolor}
\usepackage{cite}
\usepackage{amsmath,amssymb,amsfonts}
\usepackage{algorithmic}
\usepackage{graphicx}
\usepackage{textcomp}
\usepackage{siunitx} 
\usepackage{scalerel}
\usepackage{subfigure}
\usepackage{enumerate}
\usepackage{mathtools}
\usepackage{multirow}
\usepackage{amsmath}
\usepackage{enumitem}
\usepackage{makecell}
\usepackage{array}
\usepackage[]{changes}
\colorlet{Changes@Color}{blue}
\usepackage{booktabs}
\def\BibTeX{{\rm B\kern-.05em{\sc i\kern-.025em b}\kern-.08em
    T\kern-.1667em\lower.7ex\hbox{E}\kern-.125emX}}
\begin{document}

\title{Multiple-Frequency-Bands Channel Characterization for In-vehicle Wireless Networks}

\author{Mengting Li, Yifa Li, Qiyu Zeng, Kim Olesen, Fengchun Zhang and Wei Fan

\thanks{This work has been submitted to the IEEE for possible publication. Copyright may be transferred without notice, after which this version may no longer be accessible.}
 
\thanks{Mengting Li, Yifa Li, Qiyu Zeng, Kim Olesen and Fengchun Zhang are with the Antenna Propagation and Millimeter-wave Systems (APMS)
section, Aalborg University, Aalborg,  Denmark.  }

\thanks{Wei Fan is with the National Mobile Communications Research Laboratory, School of Information Science and Engineering, Southeast University, Nanjing, China.}

\thanks{Corresponding author: Wei Fan (Email: weifan@seu.edu.cn).}}


\maketitle

\begin{abstract}
In-vehicle wireless networks are crucial for advancing smart transportation systems and enhancing interaction among vehicles and their occupants. However, there are limited studies in the current state of the art that investigate the in-vehicle channel characteristics in multiple frequency bands. In this paper, we present measurement campaigns conducted in a van and a car across below 7 GHz, millimeter-wave (mmWave), and sub-Terahertz (Sub-THz) bands. These campaigns aim to compare the channel characteristics for in-vehicle scenarios across various frequency bands. Channel impulse responses (CIRs) were measured at various locations distributed across the engine compartment of both the van and car. 
 The CIR results reveal a high similarity in the delay properties between frequency bands below 7GHz and mmWave bands for the measurements in the engine bay. Sparse channels can be observed at Sub-THz bands in the engine bay scenarios. Channel spatial profiles in the passenger cabin of both the van and car are obtained by the directional scan sounding scheme for three bands. We compare the power angle delay profiles (PADPs) measured at different frequency bands in two line of sight (LOS) scenarios and one non-LOS (NLOS) scenario. Some major \added{multipath components (MPCs)} can be identified in all frequency bands and their trajectories are traced based on the geometry of the vehicles. The angular spread of arrival is also calculated for three scenarios. The analysis of channel characteristics in this paper can enhance our understanding of in-vehicle channels and foster the evolution of in-vehicle wireless networks.
\end{abstract}


\section{Introduction}
\IEEEPARstart{V}{ehicle} 
-to-everything (V2X) systems are envisioned to achieve broad connectivity between vehicles and any relevant entity such as infrastructures, wireless devices, pedestrians etc. for the fifth generation (5G) communication systems. Extensive channel measurements have been conducted to explore the channel characteristics in these scenarios \cite{li2020measurements,bai20213d,yang2023dynamic,yusuf2020experimental,doone2018pedestrian,yang2021non,rashdan2018vehicle,jiang2023hybrid}. 
\added{As one of the key driving factors for the ambitious sixth generation (6G) visions, in-X subnetworks have attracted extensive attention from both the academia and industries\cite{adeogun2020towards,gautam2023cooperative,berardinelli2021extreme,du2022multi}. While the ‘X’ represents entity where the subnetwork is installed, e.g., a robot, a vehicle, or even a human body, ‘subnetworks’ indicate that these cells are able to work in a stand-alone style but may also be connected to the 6G wide area network. These short-range, hyper-dense, highly specialized subnetworks aim to unleash the industry 4.0 factory where wireless replaces cables for achieving extremely reliable and low latency cycles in demanding scenarios. In-vehicle subnetworks which under the architecture of in-X subnetworks are the focus of this work.}

\added{\textit{A. Motivation}}

 \added{In-vehicle communications are at present supported by cable-based communication protocols such as the controller area network (CAN) bus \cite{ran2010design}, automotive Ethernet \cite{hank2013automotive}, which assist in connecting sensors, actuators and electronic control units (ECUs). A car typically contains more than a mile of wiring, making it the third-heaviest system in the vehicle. With growing number of automotive applications, the number of sensors, e.g., radars, lidar and central control unit has largely increased in vehicles, resulting in a more challenging cable management. The transition from cable-based communication to wireless networks in vehicles offers greater flexibility in sensor placement, enhances vehicle sustainability by reducing weight, and eliminates concerns related to cable aging. In-vehicle subnetworks face several significant challenges. An in-vehicle subnetwork installed in a moving vehicle on a congested road could face various interference from both the internal and external sources including jamming attacks which can disrupt the communication link quality. Power consumption consideration, involvement of additional transceiver designs and the establishment of industry standards may also pose further challenges.}

\added{ Radio channel modeling is essential and serves as the foundation for developing new communication systems. The sensors and actuators could be distributed all over the vehicles while the control units are typically in the engine bay, passenger cabins, under the passenger floorboard, under the seats, or in the luggage areas. Channel characterization in these scenarios is fundamental for developing in-vehicle wireless networks. However, fewer in-vehicle channel measurements have been reported in state of the arts compared to measurements conducted outside the vehicles.}

 \textit{\added{B. Use Cases and Spectrum Selection}}

 \added{In-vehicle subnetworks aim to use short range wireless networks for supporting use cases that demand extreme performance. The potential use cases could be engine control, power steering, antilock braking system (ABS), etc. The general requirements for in-vehicle subnetworks supporting these use cases are summarized in Table \ref{tab:use_cases}. High-criticality data flows demand latencies as low as \SI{54}{\us}, with reliability exceeding six nines. Additionally, critical traffic may coexist with high data rate applications, such as those required by advanced driver-assistance systems (ADAS). As parts of the 6G radio system, in-vehicle subnetworks may also have the capability of accessing the wide area network with medium or non-critical flows for assisting vehicular-to-vehicular services such as lane changes, intersection movement assist etc. Support for the extreme communication requirements of in-vehicle subnetworks calls for the use of a large spectrum band.}

\added{At present, there are no established standards for spectrum selection in in-vehicle subnetworks. The frequency bands below 6 GHz have the favorable propagation conditions compared to millimeter wave (mmWave) and higher frequencies. However, bands between 1.7 GHz and 4.7 GHz have already been overcrowded, which makes it challenging to support the most demanding applications. mmWave and Sub-Terahertz (Sub-THz) bands, with large spectrum resources may have the capability of supporting challenging low latency applications whereas the high propagation and scattering losses make these frequencies not favorable for non-line-of-sight (NLOS) scenarios. 5 and 6 GHz bands are less populated than other Sub-6 GHz bands and could be possible options for in-vehicle subnetworks. In addition, more studies of in-vehicle channels at mmWave and Sub-Thz are needed to assess their potential in meeting extreme performance requirements. Therefore, spectrum around 5 GHz, 26, 28 GHz and 100 GHz are selected to be the focus of our work.}

 \begin{table}
  \begin{center}
    \caption{\added{General Requirements for In-vehicle Wireless Subnetworks.}}
    \label{tab:use_cases}
    \setlength{\tabcolsep}{1.5pt}
\renewcommand\arraystretch{1.2}
    \begin{tabular}{|c|c|}
    \hline
      \textbf{Parameters} & \textbf{Value} \\
    \hline
      Maximum range & $\sim 10 m $ \\
    \hline
      Data rate per link & \makecell[tc]{$<$ 10 Mbps (for control)\\$<$10 Gbps (for ADAS sensors)}\\
   \hline
      Minimum latency & $\sim$ \SI{54}{\us}\\
   \hline
     \makecell[tc]{Communication\\service availability} & 99.9999\% to 99.999999\% \\
       \hline
    \end{tabular}
  \end{center}
\end{table}

\textit{\added{C. State-of-the-art}}

In the state-of-the-arts, most of in-vehicle channel measurements were conducted in the ultra wideband (UWB) frequencies (3.1-10.6 GHz) due to their abundant spectrum resources. 
The channels in the engine bay \cite{demir2013engine}, passenger cabins \cite{schack2008measurements,blumenstein2014vehicle} and trunk \cite{schack2009uwb,blumenstein2015measurements} of the vehicles were measured at UWB bands. The pathloss and the delay properties were analyzed based on the measured data. 
In \cite{blumenstein2014vehicle}, the authors focused on characterizing the spatial stationarity of channels measured at UWB band.      
As mmWave bands are extensively explored in 5G, in-vehicle channel measurements in mmWave bands and comparisons with lower frequency bands have been conducted. In \cite{schack2010comparison}, the pathloss and root mean square (RMS) delay spread results were compared between 5-8.5 GHz and 67-70.5 GHz channels. A more comprehensive comparison of in-vehicle channels in terms of resolvable clusters, RMS delay and spatial stationarity between UWB and 60 GHz was presented in \cite{blumenstein2017vehicle}. In \cite{yamakawa2021experimental}, the multipath parameters of dominant propagation paths were extracted from channel spatial measurements in the passenger cabin at 60 GHz. 
The authors in \cite{fu2019thz} conducted channel measurements within a metal enclosure of desktop size in the range of 280–320 GHz to explore the viability of wireless chip-to-chip communication in Sub-THz bands. \added{Comparisons of channel characteristics between 30 GHz, 140 GHz and 300 GHz have been investigated in \cite{cheng2017comparison,cheng2018study} for short range communication. The pathloss and diffraction properties at these frequencies are characterized using appropriate channel models.}
Exploring in-vehicle channels at Sub-THz is also valuable for addressing the demanding use cases of in-vehicle wireless networks.

\textit{\added{D. Contribution}}

To date, there have been no reported in-vehicle channel measurements conducted in sub-THz and THz frequency bands.
Furthermore, most of the previous in-vehicle channel measurements have concentrated on scenarios within passenger cabins and trunks. Only a few have been conducted in the engine bay, despite this scenario being crucial for wireless networks between sensors and control units. Additionally, channel spatial profiles are vital for multi-antenna systems. However, in-vehicle channel spatial profiles in high-frequency bands remain largely unexplored. Our study in this paper aims to address these gaps, and the main contributions are listed below:

\begin{itemize}
 \item 

Extensive channel measurements have been conducted in both a van and a car across frequency bands below 7 GHz, mmWave and Sub-Thz bands. These measurements enable the exploration of frequency independence for in-vehicle channels.

\item Channel measurements in the engine bay of both the van and car have been conducted for multiple bands. We analyze and compare the acquired channel impulse responses (CIRs), focusing on the pathloss, and RMS delay spread across these three frequency bands.
 \item Channel spatial profiles of two Line-of-sight (LOS) scenarios and one non LOS (NLOS) scenario for above mentioned three bands have been measured by the directional scan sounding scheme (DSS). We discuss the identified major multipath components (MPCs) obtained from the measured power angle delay profile (PADP) and calculate the angular spread of arrival for each scenario. 
\end{itemize}

The rest of this paper is organized as follows. In Section \ref{sec:measurements}, we introduce the in-vehicle channel measurement campaigns for different scenarios. In Section \ref{sec:results_pdp}, we present exemplary CIR results and analyze the results in terms of pathloss, and RMS delay spread. In Section \ref{sec:results_padp}, we present the results of channel spatial measurements and analyze the spatial characteristics of the MPCs. Finally, Section \ref{sec:conclusion} draws the conclusion.


\section{In-vehicle Channel Measurements} \label{sec:measurements}
Our measurements utilized two vehicles: a van (Ford Transit Custom model) and a car (Benz B200d model). The in-vehicle channel measurements presented in this paper can be divided into two categories. \added{The first category of measurements, conducted in the engine bay, aims to capture power and delay characteristics. Spatial properties are not taken into account because, in practical applications, the engine bay scenario has limited potential for utilizing spatial dimensions due to the restricted space available for large arrays. Furthermore, it is challenging to install a rotator in the engine bay in our measurements. The second category, conducted in the passenger cabin, aims to capture channel spatial profiles. Due to the relatively long measurement times and the challenges of providing mechanical support for the antennas, the spatial measurements are limited to three scenarios with three different Tx-Rx locations. The vehicles used in our measurements are internal combustion engine (ICE) vehicles. The propagation environments inside vehicles may vary depending on factors such as vehicle type (e.g., car, truck) and brand. However, the differences between ICE and electric vehicles of similar models are not expected to be significant. Therefore, the channel characteristics captured in our measurements can be valuable for developing channel models for both electric and ICE vehicles.}

\subsection{Measurements in the Engine Bay}


\subsubsection{Measurement System}
\begin{figure*}
\centering
{\includegraphics[width=1.6\columnwidth]{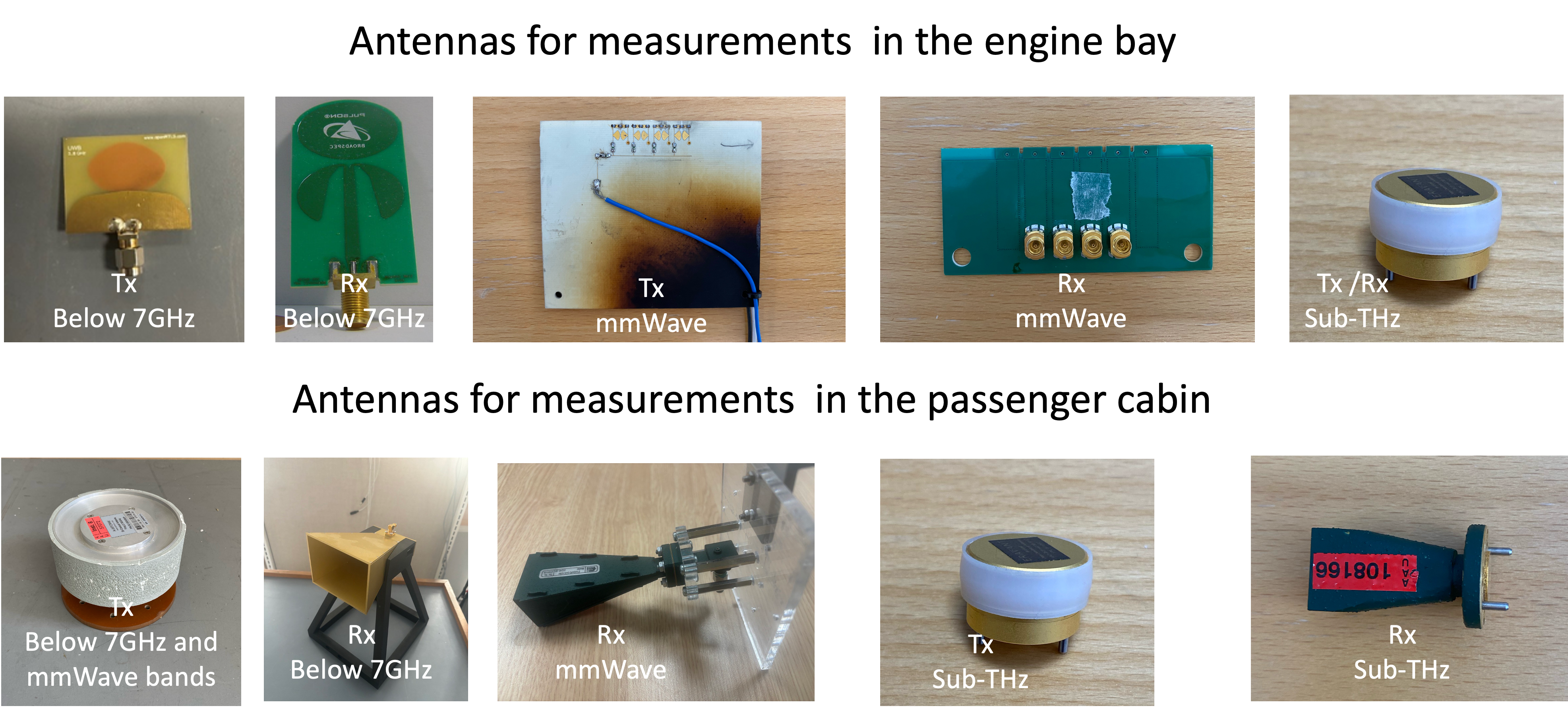}}

\caption{Photos of the antennas used in the measurements.}
\label{fig:antenna}
\end{figure*}
 The R\&S ZNA43 vector network analyzer (VNA) is used to record the channel frequency response (CFR), i.e., S21 for each paired transmitting (Tx) antenna and receiving (Rx) antenna. Note that frequency extender modules (VDI Model: WR10-VNAX) are incorporated into the VNA-based channel sounder to measure the channel data from 99 to 101.5 GHz. Back-to-back measurements were conducted to calibrate the system response before data collection. The channel measurements were conducted in three different bands. The measurement settings and antenna specifications are summarized in Table \ref{tab:pdp_mea}. The measured frequency bandwidth is 2.5 GHz for all the three bands, resulting in a 0.4 ns delay resolution. For each frequency band, 626 frequency points have been recorded and with a delay range of 250 ns allowing for a maximum measurement range of 75 m. As the frequency goes up, a smaller intermediate frequency (IF) bandwidth is set to maintain a sufficient dynamic range of the measurement system. The selection of the antennas used in the engine bay scenario could be problematic. The physical size of the antennas should be small for easy installation in the engine bay. Moreover, the antenna patterns used in three bands should be similar to exclude the effects of antenna patterns. The photos of the antennas used in channel measurements are given in Fig.\ref{fig:antenna}. Planar monopoles and biconical antennas, both with omni-directional patterns in the horizontal plane are employed for measurements in the lowest frequency and Sub-Thz bands, respectively. Due to the lack of small omni-directional antennas in the mmWave bands, endfire antennas with wide half power beamwidth (HPBW) in both horizontal and vertical planes are selected for measurements in mmWave bands. The model numbers of the planar monopoles are OpenRTLS UWB antenna (Tx) and Pulson Broad2p3c (Rx), respectively. More details of endfire antennas and biconical antennas can be referred to \cite{zhang2019radiation,zhang2021wideband} and \cite{omni}, respectively. Note that the orientation of the endfire antennas was carefully arranged to ensure the main beam of the Tx and Rx is generally aligned. The purpose is to reduce the observed channel differences caused by the antennas. All of the employed antennas are vertically polarized.

 \begin{table}
  \begin{center}
    \caption{Specifications of Measurements in the Engine Bay.}
    \label{tab:pdp_mea}
    \setlength{\tabcolsep}{1.5pt}
\renewcommand\arraystretch{1.2}
    \begin{tabular}{|c|c|c|c|}
    \hline
      \textbf{Parameters} & \textbf{Below 7 GHz} & \textbf{mmWave }& \textbf{Sub-Thz }\\
      \hline
      Frequency band [GHz] & 4-6.5 & 25-27.5 & 99-101.5\\
      \hline
     Frequency points & \multicolumn{3}{c|}{626}\\
        \hline
     Bandwidth [GHz]& \multicolumn{3}{c|}{2.5}\\
      \hline
     Transmitted power [dBm] & \multicolumn{3}{c|}{10}\\
     \hline
     IF bandwidth [Hz] & 1000 & 500 & 100\\
       \hline
    \makecell[tc]{Tx antenna\\ type} & \makecell[tc]{Planar \\monopole} & \makecell[tc]{endfire \\antenna} & \makecell[tc]{biconical \\antenna}\\
       \hline
   \makecell[tc]{Tx antenna gain [dBi]}  & 0.82 & 3.8 & 2\\
       \hline
    \makecell[tc]{Tx antenna \\HPBW [deg]}& \makecell[tc]{elevation:65\\ azimuth: omni\\directional} &\makecell[tc]{elevation: 60\\ azimuth: 90} & \makecell[tc]{elevation: 30\\ azimuth: omni\\directional} \\
       \hline
     \makecell[tc]{Rx antenna\\ type} & \makecell[tc]{Planar \\monopole} & \makecell[tc]{endfire \\antenna}& \makecell[tc]{biconical \\antenna}\\
       \hline
    \makecell[tc]{Rx antenna gain [dBi]}& 3 & -0.3 & 2\\
       \hline
   \makecell[tc]{Rx antenna \\HPBW [deg]} & \makecell[tc]{elevation: 70\\ azimuth: omni\\directional}  & \makecell[tc]{elevation: 70\\ azimuth: 100} & \makecell[tc]{elevation: 30\\ azimuth: omni\\directional}\\
       \hline
    \end{tabular}
  \end{center}
\end{table}

 \subsubsection{Measurement Scenarios}
\begin{figure}
\centering
\subfigure []
{\includegraphics[width=0.9\columnwidth]{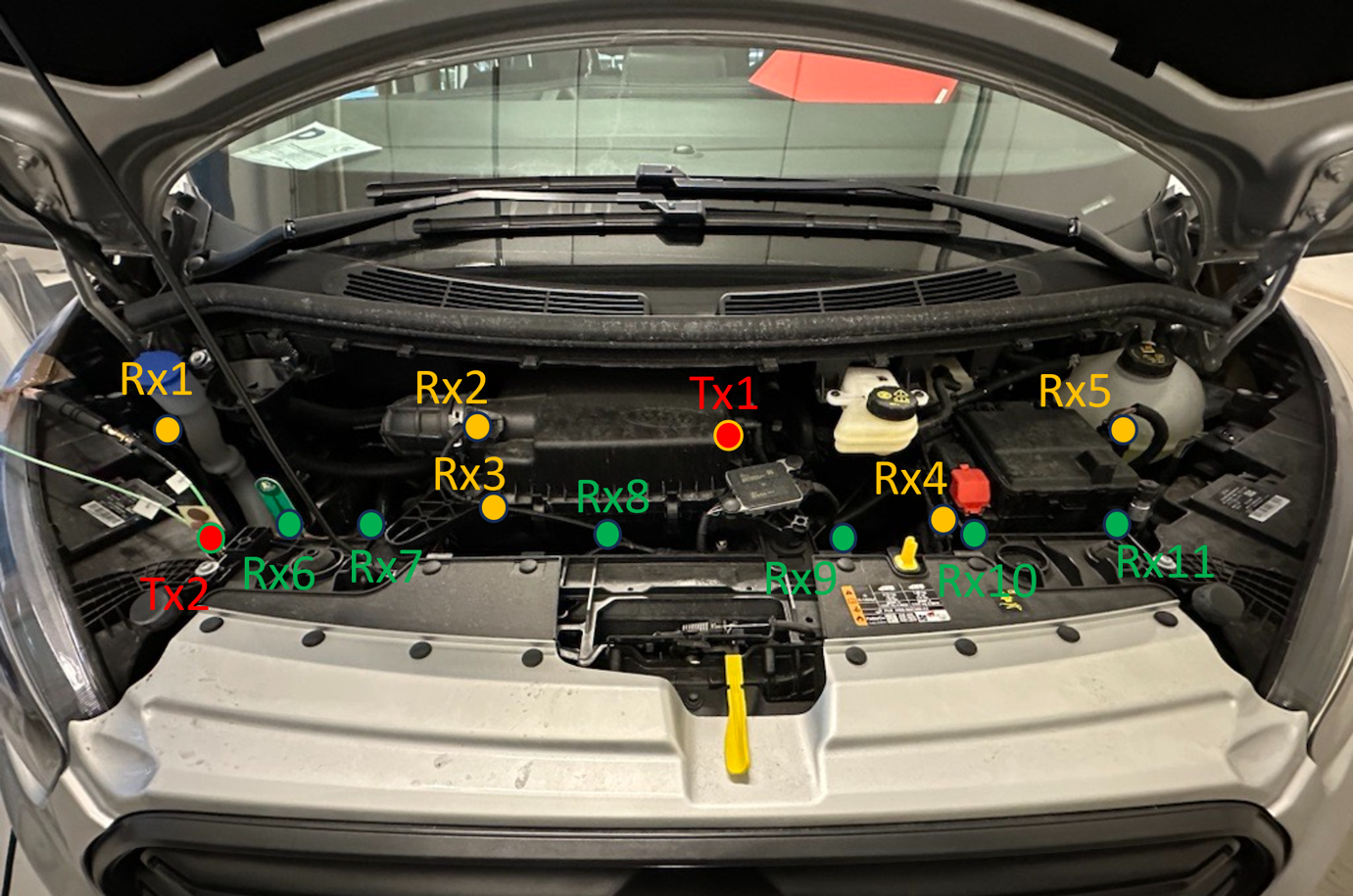}}
\subfigure []
{\includegraphics[width=0.9\columnwidth]{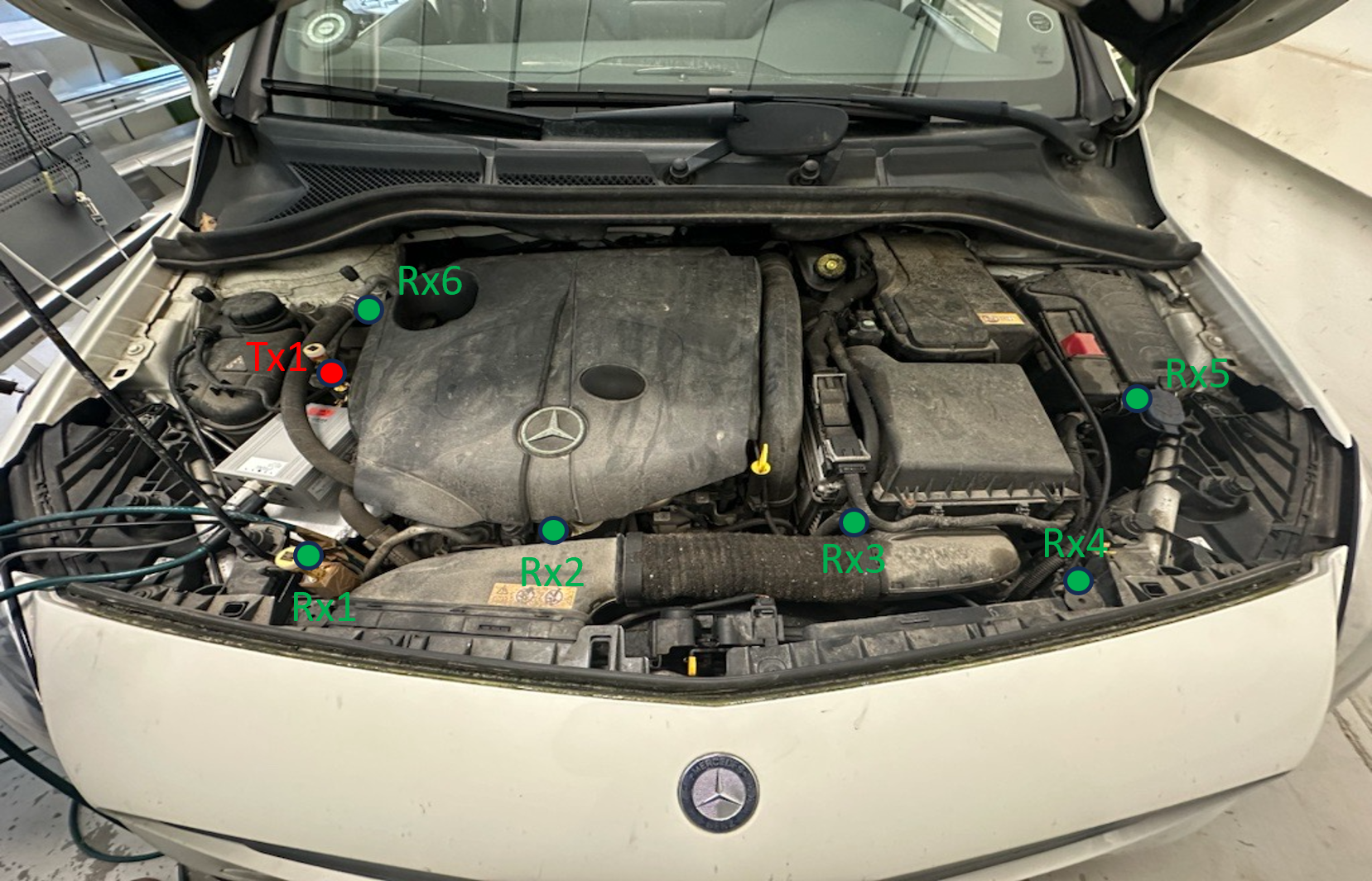}}

\caption{Measurement locations in the engine bay of the (a) van and (b) the car, respectively.}
\label{fig:pdp_map}
\end{figure}

The Tx and Rx locations within the engine of the van are illustrated in Fig \ref{fig:pdp_map} (a). Rx 1-5 are paired with Tx 1, while Rx 6-11 are paired with Tx 2. The Rx locations are distributed among the whole engine bay area.  
Due to the limited space in the engine bay area of the car, fewer samples were measured as shown in Fig \ref{fig:pdp_map} (b). The front cover of the vehicles was closed during the measurements in the engine bay to approximate the realistic scenarios. 

\subsection{Measurements in the Passenger Cabin}
\subsubsection{Measurement System}
\begin{figure}
\centering
{\includegraphics[width=1\columnwidth]{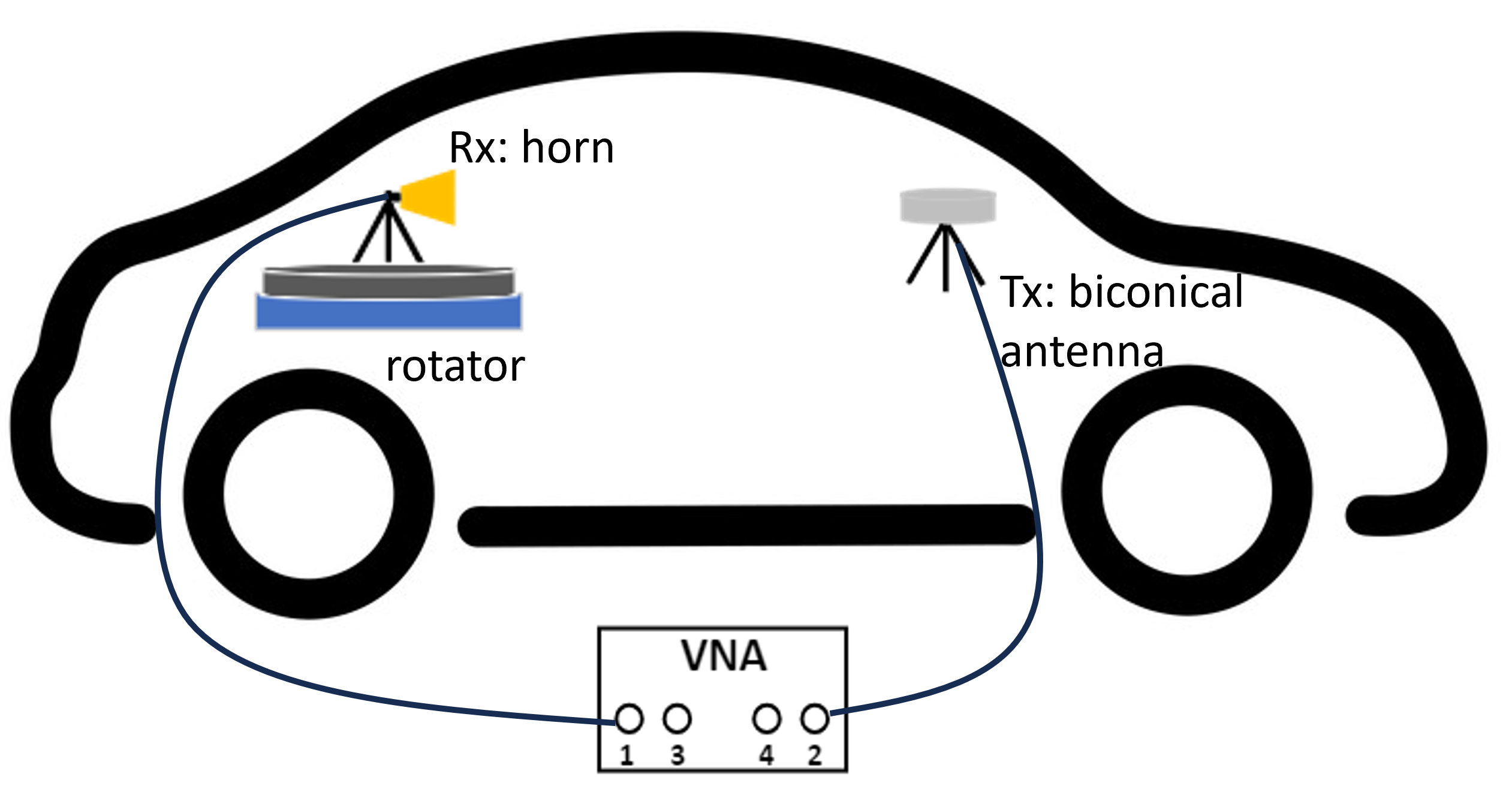}}

\caption{The diagram of the measurement system for measurements in the passenger cabin.}
\label{fig:sys_spatial}
\end{figure}
Spatial information of the MPCs is essential for wireless network employing multi-antenna systems to evaluate and improve the link budget and data rate. DSS scheme \cite{bengtson2022long,lyu2023sub} was used to obtain the channel spatial profiles for different scenarios inside the vehicle. The diagram of the measurement system is given in Fig. \ref{fig:sys_spatial}. 
Two biconical antennas with omnidirectional patterns in the azimuth plane were employed as Tx antennas. The first biconical antenna \cite{biconical} operates in both the lower frequency and mmWave bands, while the second operates in Sub-Thz bands \cite{omni}. Additionally, three distinct horn antennas \cite{horn} were chosen as Rx antennas for the three bands. The photos of the utilized antennas are shown in Fig. \ref{fig:antenna}. The Rx antenna was mounted on a mechanical rotator, enabling it to capture signals from various directions as the rotator rotates. The rotation step is the same i.e., 10$^\circ$ for all the frequency bands for ease of implementation. The VNA recorded 501 frequency points for each band at each rotation angle. Detailed measurement settings and antenna specifications can be found in Table \ref{tab:spatial_mea}.

 \begin{table}
  \begin{center}
    \caption{Summary of Channel Spatial Measurements.}
    \label{tab:spatial_mea}
    \setlength{\tabcolsep}{1.5pt}
\renewcommand\arraystretch{1.2}
    \begin{tabular}{|c|c|c|c|}
    \hline
      \textbf{Parameters} & \textbf{Below 7 GHz} & \textbf{mmWave }& \textbf{Sub-Thz }\\
      \hline
      Frequency band [GHz] & 4.4-6.4 & 28-30 & 99-101\\
       \hline
      Bandwidth [GHz] & \multicolumn{3}{c|}{2}\\
      \hline
     Frequency points &\multicolumn{3}{c|}{501}\\
      \hline
     IF bandwidth [Hz] & 1000 & 500 & 100\\
       \hline
     Rotation step [deg] & 10 & 10 & 10\\
       \hline
    \makecell[tc]{Tx antenna\\ type} &  \multicolumn{3}{c|}{biconical antenna}\\
       \hline
   \makecell[tc]{Tx antenna gain [dBi]}  & {4}& {4} & 5\\
       \hline
     \makecell[tc]{Rx antenna\\ type} &  \multicolumn{3}{c|}{horn antenna}\\
       \hline
    \makecell[tc]{Rx antenna gain [dBi]}& 10 & 20 & 20\\
       \hline
   \makecell[tc]{Rx antenna \\HPBW [deg]} & \makecell[tc]{azimuth: 38} & \makecell[tc]{azimuth: 17} & \makecell[tc]{azimuth: 16}\\
       \hline
   Scenario & \makecell[tc]{van:LOS,NLOS \\ car: LOS} & \makecell[tc]{van:LOS,NLOS \\ car: LOS} & \makecell[tc]{van:LOS\\ car: LOS}\\
       \hline
    \end{tabular}
  \end{center}
\end{table}

\subsubsection{Measurement Scenarios}
\begin{figure}
\centering
\subfigure []
{\includegraphics[width=0.8\columnwidth]{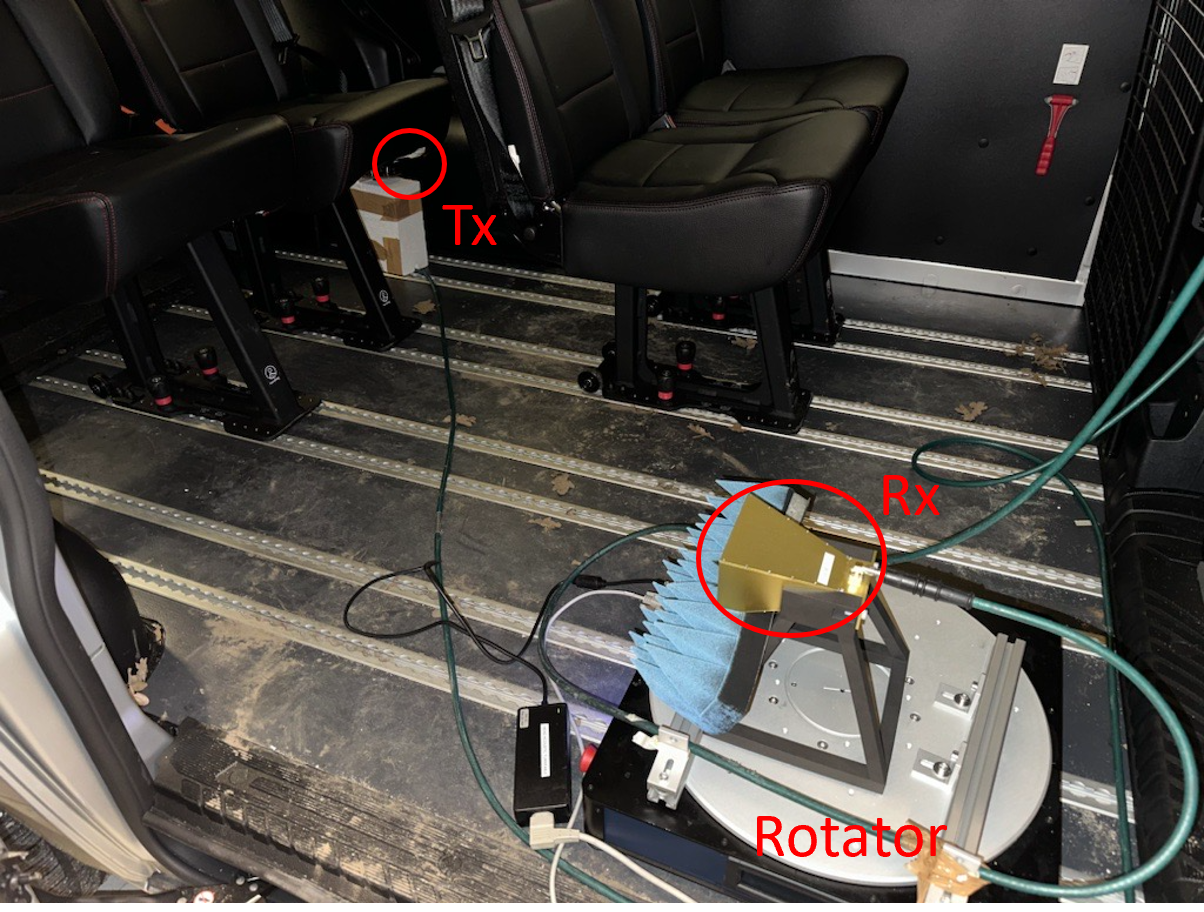}}
\subfigure []
{\includegraphics[width=0.8\columnwidth]{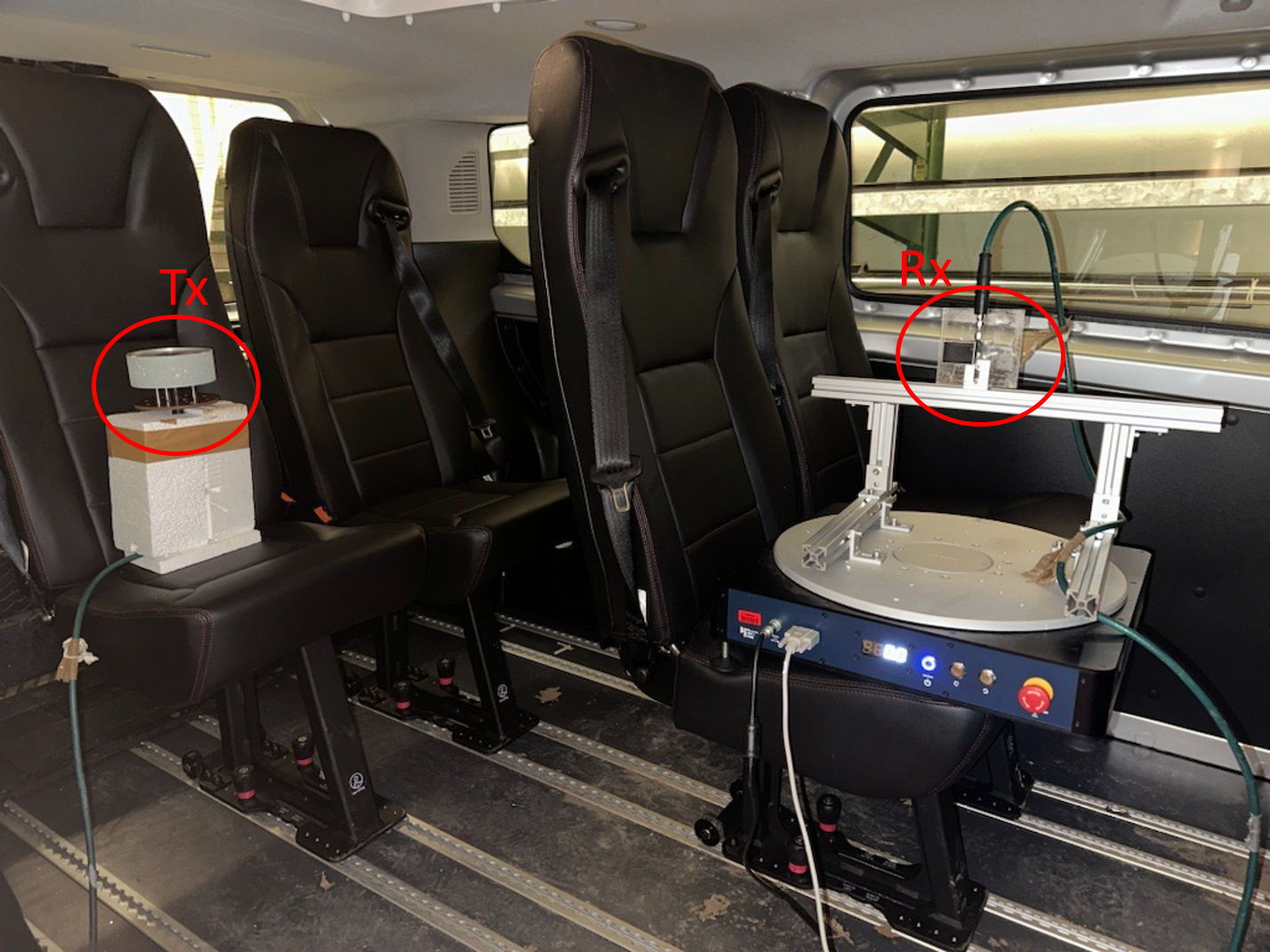}}
\subfigure []
{\includegraphics[width=0.8\columnwidth]{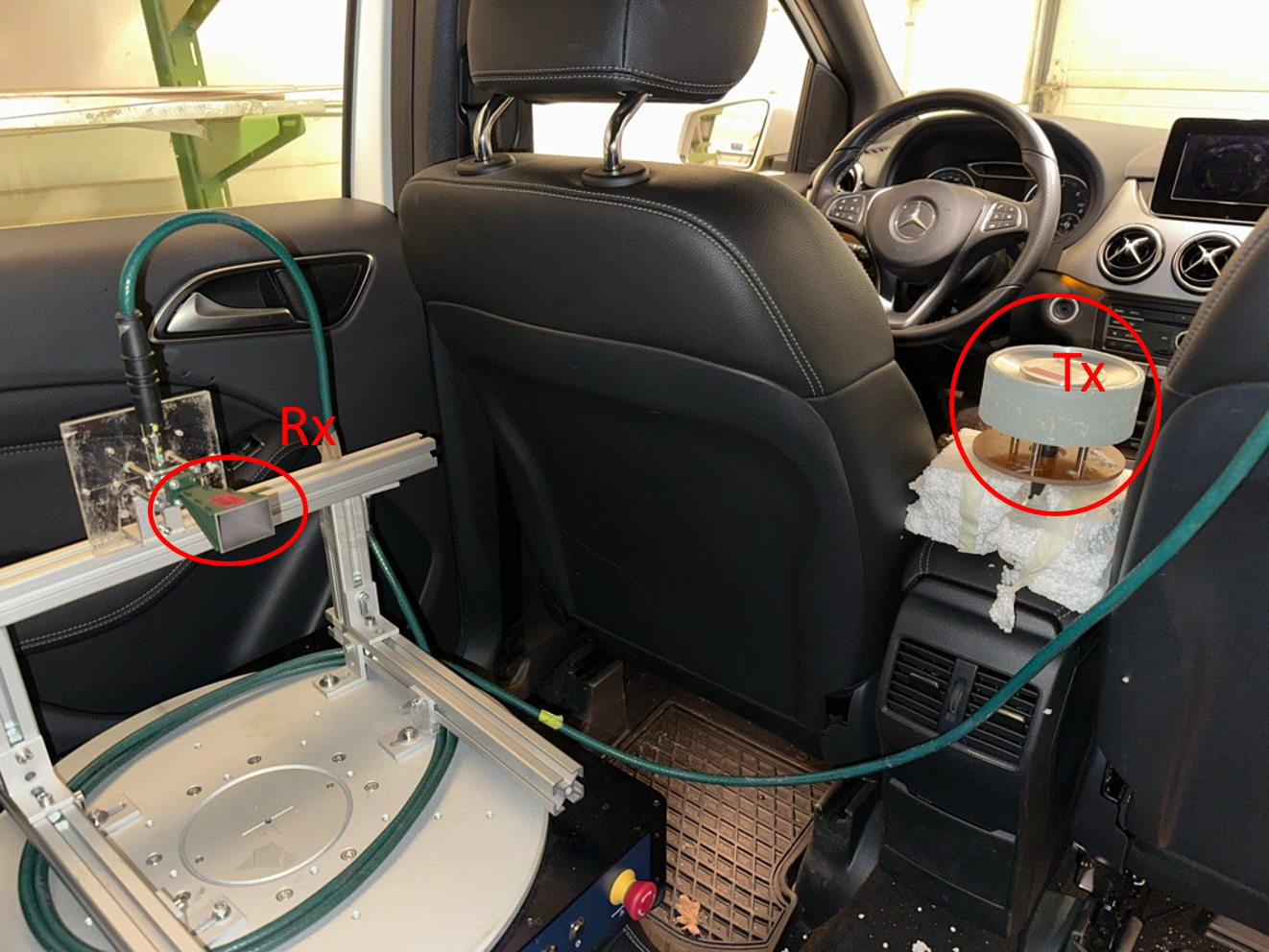}}
\caption{Photos of the channel spatial measurements in (a) Scenario 1, (b) Scenario 2 and (c) Scenario 3.}
\label{fig:scenarios_spatial}
\end{figure}
The channel spatial measurements include three distinct scenarios: Scenarios 1 and 2 were conducted within the van, while Scenario 3 was measured in the car. In Scenario 1, the Tx and Rx antennas were positioned on the floor of the passenger cabin, offering a LOS path. Scenario 2 represented a NLOS scenario, where the Tx and Rx antennas were placed on the passenger seats of the van. 
The interior space of the car is much smaller compared to that of the van, thus only a LOS scenario (Scenario 3), was considered within the car. The specific placements of Tx and Rx antennas for these scenarios are illustrated in Fig. \ref{fig:scenarios_spatial}. Channel spatial profiles were measured for three frequency bands in two LOS scenarios, while only mmWave and frequencies below 7 GHz were measured in the NLOS scenario due to the limited dynamic range in Sub-Thz bands.

\subsubsection{Data Post-processing}

The measurement environments are assumed to be static environments. The power of the CIR can be obtained by
    \begin{equation}
        P(\tau) = {\left\vert \mathcal{F}^{-1}[W(f) \cdot H(f)] \right\vert} ^2,
    \end{equation}

where $\mathcal{F}^{-1}$ denotes the inverse Fourier transform. $H(f)$ and $W(f)$ represents the measured CFR and normalized Hanning window function, respectively. 
It is challenging to detect the propagation paths from the measured CIR in a constrained space with many multiple-bounce reflections, such as the engine bay scenario. Therefore, the CIR results at all delay bins above the noise threshold will be considered when calculating the \added{delay spread} and pathloss.

  \begin{equation}
      \tau_{\mathrm{DS}} = \sqrt{\frac{\sum_{n=1}^{N}(\tau_n - \overline{\tau})^2 P(\tau_n)}{\sum_{n=1}^{N}P(\tau_n)}},
    \end{equation}
where
\begin{equation}
  \added{  \overline{\tau} = \frac{\sum_{n=1}^{N}\tau_{n}P(\tau_n)}{\sum_{n=1}^{N}P(\tau_n)}.}
\end{equation}
$N$ is the length of the delay bins. $\tau_n$ and $P(\tau_n)$ are the delay and path power of the $n$-th delay bin, respectively. $\overline{\tau}$ represents the mean delay of the measured channel.

The PADP obtained in the channel spatial measurements can be obtained by

  \begin{equation}
        PADP(\phi_{i},\tau) = {\left\vert \mathcal{F}^{-1}[\added{{W}(f)} \cdot \widetilde{H}(\phi_{i},f)] \right\vert} ^2,
    \end{equation}
where $\widetilde{H}(\phi_{i},f)$ is the measured CFR at the $i$-th rotation angle. In this case, the MPCs and their corresponding parameters are extracted by performing a 2D peak detection on the obtained PADP \cite{haneda2015statistical}. We attach more importance on the angle information of the MPCs for channel spatial measurements. Angular spread which evaluates the degree of dispersion in the angular domain can be expressed as
 \begin{equation}
      \sigma_\mathrm{AS} = \sqrt{\frac{\sum_{m=1}^{M}(\varphi_m - \mu_\mathrm{APS})^2 {{|\alpha_m|}^2}}{\sum_{m=1}^{M}{{|\alpha_m|}^2}}},
    \end{equation}
where
\begin{equation}
    \mu_\mathrm{APS} = \frac{\sum_{m=1}^{M}\varphi_m {{|\alpha_m|}^2}}{\sum_{m=1}^{M}{{|\alpha_m|}^2}}.
\end{equation}
$M$ is the total number of the identified multipath obtained from the PADP. $\varphi_m$ and $\alpha_m$ represent the angle and the complex amplitude of the $m$-th path, respectively.

\section{Power and Delay Characteristics}\label{sec:results_pdp}

\begin{figure}
\centering
\subfigure []
{\includegraphics[width=1\columnwidth]{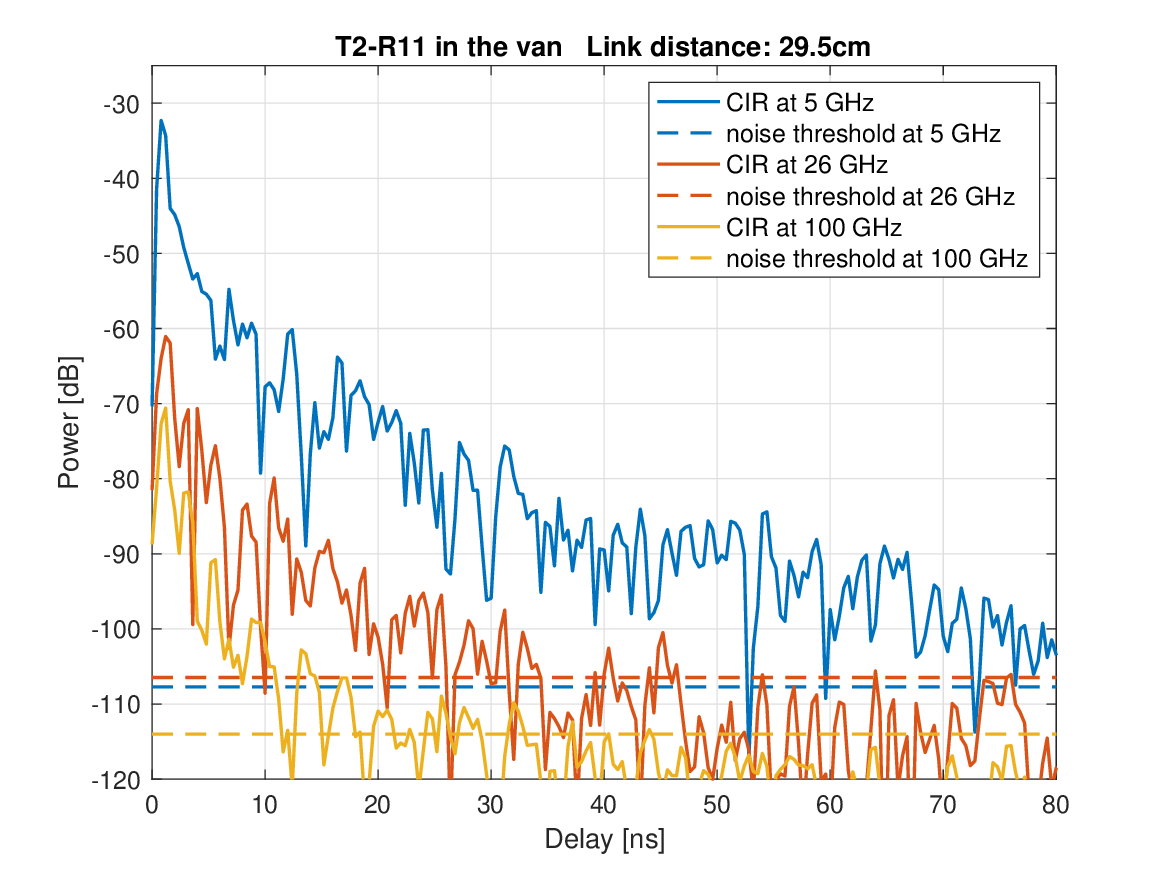}}
\subfigure []
{\includegraphics[width=1\columnwidth]{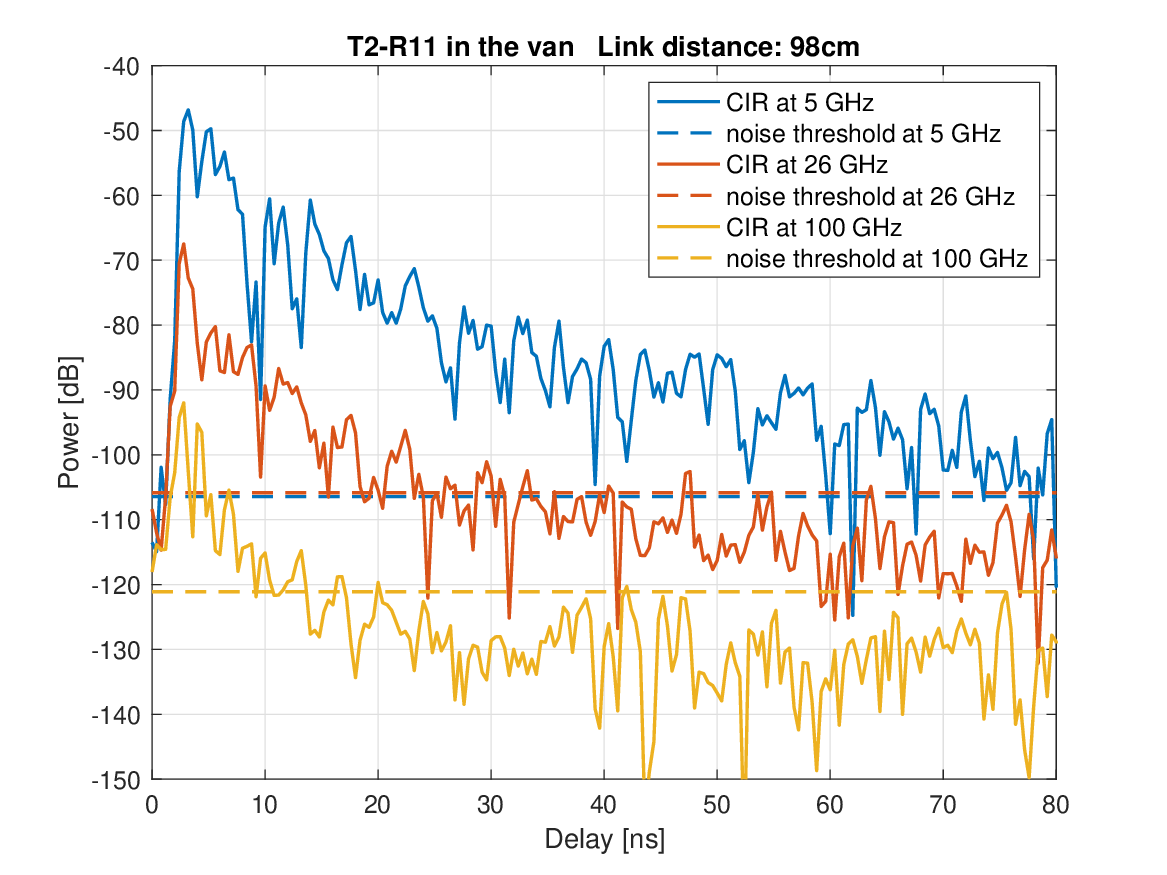}}
\subfigure []
{\includegraphics[width=1\columnwidth]{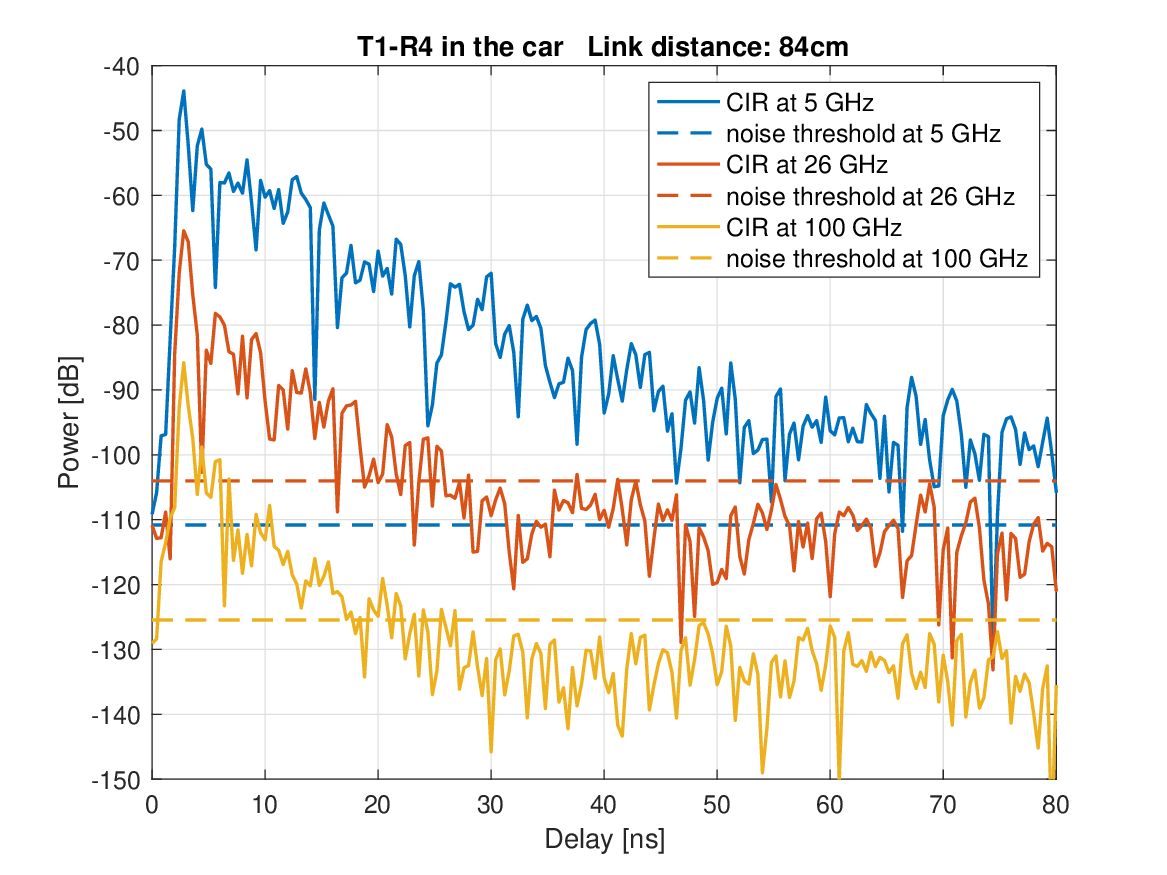}}
\caption{\added{Measured CIRs for (a) Tx1-Rx3 in the van, (b) Tx2-Rx11 in the van and (c) Tx1-Rx4 in the car, respectively.}}
\label{fig:pdp}
\end{figure}

 \added{Fig. \ref{fig:pdp} (a), (b), and (c) present three examples of CIR measurement results with link distances of 29.5 cm (Tx 1-Rx 3), 98.0 cm (Tx 2-Rx 11) in the van, and 84.0 cm (Tx 1-Rx 4) in the car, respectively. The noise threshold is determined to be 5 dB above the noise floor, i.e., the average noise power in the noise-only region.} Note that the antenna gain is embedded in the CIR results. \added{LOS path can be observed in all three scenarios and the propagation loss increases as the frequency rises. For the CIR results below 7 GHz, rich MPCs can be observed at the delay of over 80 ns whereas the MPCs mainly appear within the delay of 45 ns and 25 ns for mmWave and Sub-THz bands, respectively. Fig. \ref{fig:pdp} (a) shows the CIR results measured in the van with a short link distance of 29.5 cm and the power of the LOS path is much greater than the other multipath. Comparing the CIR results at different frequencies in Fig. 5 (a), some strong MPCs can only be observed in mmWave bands. These differences could be caused by the use of antennas with different patterns for these three bands or the antenna location alignment errors during the process of changing antennas. Fig. \ref{fig:pdp} (b) is the result with a longer link distance in the engine bay of the van and the path powers drop less dramatically as the delay increases compared to Fig. \ref{fig:pdp} (a). Fig. \ref{fig:pdp} (b) and (c) show the CIR results from different vehicles with similar link distances. Comparing Fig. \ref{fig:pdp} (b) and (c), similar strong MPCs can be identified within 10 ns delay in both figures. These strong MPCs might result from the multi-bounce reflections off the metal front and side covers of the vehicles. The differences in other MPCs shown in Fig. \ref{fig:pdp} (b) and (c) could be attributed to the different structures in the engine bay of these two vehicles. The engine bay of the van has a deeper depth than that of the car and the engine components are distributed with different heights, resulting in a more complicated propagation scenario. MPCs can be scattered from the edge of the components. On the other hand, the engine bay of the car is composed of several big plastic blocks located in the similar heights. MPCs are mainly the reflections from the plastic blocks.}

\begin{figure}
\centering
\subfigure []
{\includegraphics[width=1\columnwidth]{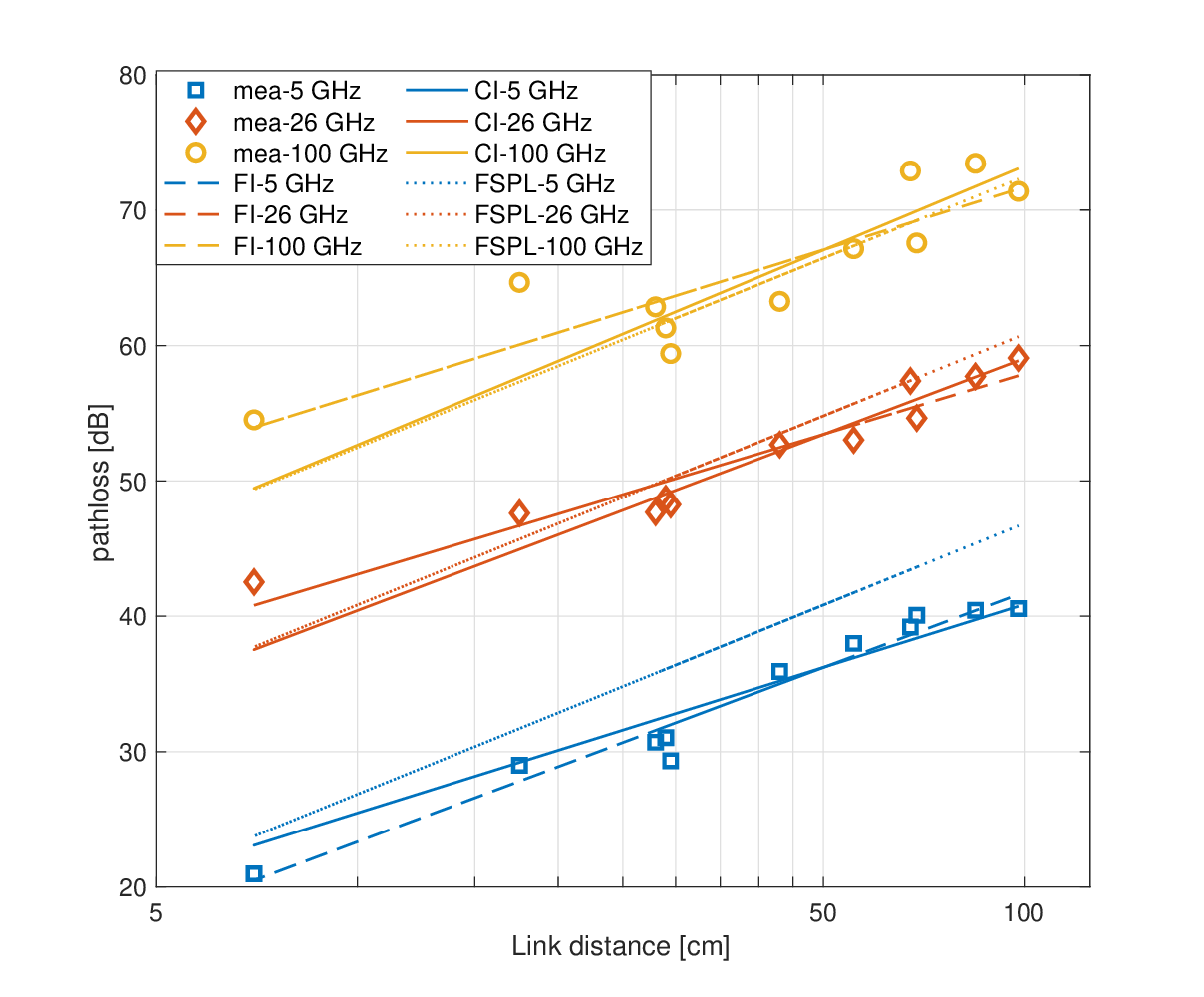}}
\subfigure []
{\includegraphics[width=1\columnwidth]{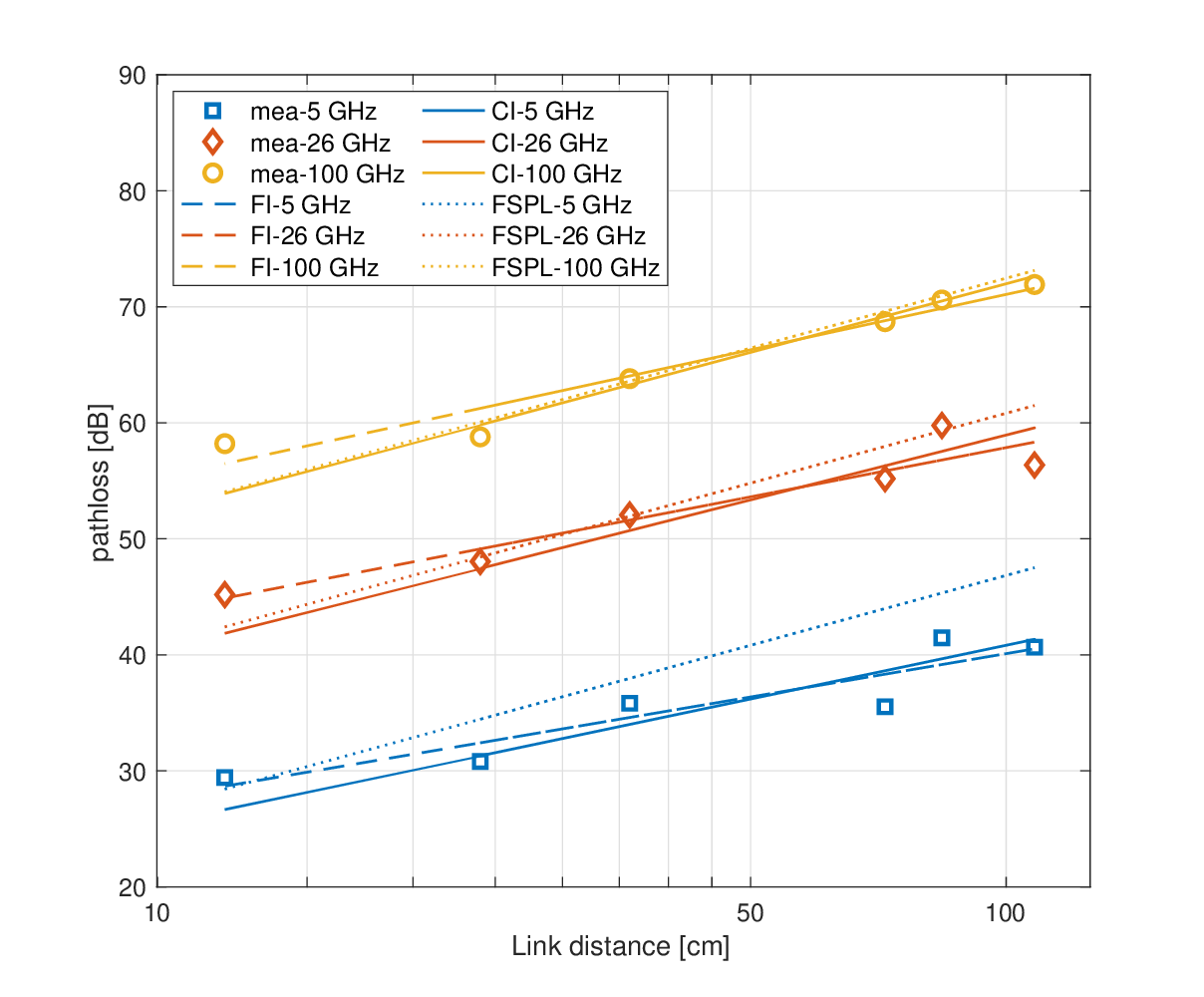}}

\caption{\added{The pathloss and fitting model results for measurements in the engine bay of (a) the van and (b) the car, respectively.}}
\label{fig:pathloss_model}
\end{figure}

\added{Single-frequency floating-intercept (FI) and the single-frequency close-in (CI) free space reference distance model are used here to fit the measured pathloss. The FI-model used in the WINNER II and the 3rd Generation Partnership Project (3GPP) is defined as 
\begin{equation}
     \mathrm{PL}^\mathrm{FI} (d) = \alpha + 10\beta\mathrm{log}_{10}(\frac{d}{d_0}) + X^{\mathrm{FI}}_\sigma (d \geq d_0),
\end{equation}
 where $d$ represents the link distance and $d_0 = 5 cm$ is the selected reference distance. $\alpha$ is the floating intercept in decibels, $\beta$ is the pathloss exponent and $X^{\mathrm{FI}}_\sigma$ is a log-normal random variable with zero mean and standard deviation $\sigma$. The CI model with fewer parameters is defined as
\begin{equation}
     \mathrm{PL}^\mathrm{CI} (d) = \mathrm{PL}(d_0) + 10\beta\mathrm{log}_{10}(\frac{d}{d_0}) + X^{\mathrm{CI}}_\sigma (d \geq d_0),
\end{equation}
 Fig. \ref{fig:pathloss_model} (a) and (b) show the results of measured pathloss, theoretical free space pathloss (FSPL) and FI, CI pathloss models for measurements in the engine bay of the van and the car, respectively. The link distance ranges from 8.5 cm to 98.0 cm for measurements in the van and from 12.0 cm to 108.0 cm for measurements in the car. Antenna gains of both Tx and Rx antennas are removed when calculating the pathloss. Based on the equations shown above, the model parameters estimated by least-squares linear fitting are shown in Table \ref{tab:pathloss_model}. }
 
 \added{For both the van and car scenarios, $\sigma$ of the FI models is always lower than that of the CI models across all frequency bands, indicating a better fitting accuracy of FI pathloss models. Note that since we do not have perfect omni-directional antennas and cannot accurately align the heights of the Tx and Rx antennas, additional errors might be introduced when de-embedding the antenna gains. FI pathloss models are more flexible with the de-embedding and measurement errors. The measured pathloss in mmWave and Sub-THz bands at several Rx locations in both van and car is higher than the corresponding FSPL. The reason could be that we overestimate the antenna gains in these scenarios. Moreover, the power of the LOS path may be diminished by scattering paths with similar delays or by destructive interference from scatterers encroaching into the Fresnel zones. Pathloss exponents smaller than the theoretical FSPL exponent of 2 are observed in all the FI models and CI models. The engine bay of the vehicle, with a metal cover above and plastic blocks or components below, forms a narrow “tunnel” in between, creating a highly reflective environment that leads to small path loss exponents.}   

{ \begin{table}
  \begin{center}
    \caption{\added{Parameters of the pathloss models.}}
    \label{tab:pathloss_model}
    \setlength{\tabcolsep}{1.2pt}
\renewcommand\arraystretch{1.2}
    \begin{tabular}{|c|c|c|c|}
    \hline
    \textbf{Scenario} & \multicolumn{3}{c|}{\textbf{Model Parameter Values}} \\
      \cline{2-4}
                             & 5 GHz & 26 GHz & 100 GHz \\ \hline
           \makecell[tc]{Engine bay\\(van)}  
                             & \makecell[tl]{FI model:\\$\alpha=17.79$, $\beta=$ \\1.84, $\sigma=1.16$\\CI model:\\ $\mathrm{PL}_(d_0)=20.82$,\\$\beta=1.54$,$\sigma=1.52$} 
                              & \makecell[tl]{FI model:\\$\alpha=38.63$, $\beta=$ \\1.48, $\sigma=1.39$\\CI model:\\ $\mathrm{PL}_(d_0)=34.80$,\\$\beta=1.86$,$\sigma=1.86$} 
                              & \makecell[tl]{FI model:\\$\alpha=51.7$, $\beta=$ \\1.54, $\sigma=2.65$\\CI model:\\ $\mathrm{PL}_(d_0)=46.44$,\\$\beta=2.06$,$\sigma=3.24$} \\ \hline
               \makecell[tc]{Engine bay\\(car)}  
                             & \makecell[tl]{FI model:\\$\alpha=23.97$, $\beta=$ \\1.24, $\sigma=1.72$\\CI model:\\ $\mathrm{PL}_(d_0)=20.82$,\\$\beta=1.54$,$\sigma=2.02$} 
                              & \makecell[tl]{FI model:\\$\alpha=39.53$, $\beta=$ \\1.41, $\sigma=1.56$\\CI model:\\ $\mathrm{PL}_(d_0)=34.80$,\\$\beta=1.86$,$\sigma=2.23$} 
                              & \makecell[tl]{FI model:\\$\alpha=50.45$, $\beta=$ \\1.58, $\sigma=1.27$\\CI model:\\ $\mathrm{PL}_(d_0)=46.44$,\\$\beta=1.96$,$\sigma=1.85$}                    
           \\
       \hline
    \end{tabular}
  \end{center}
\end{table}}

\begin{figure}
\centering
\subfigure []
{\includegraphics[width=1\columnwidth]{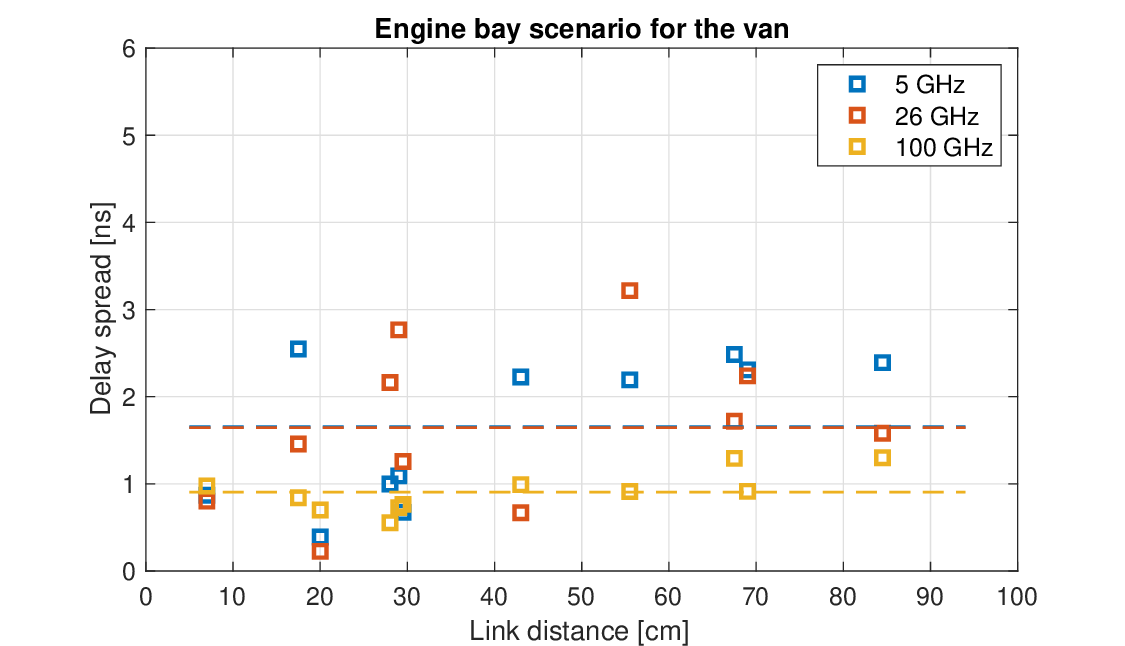}}
\subfigure []
{\includegraphics[width=1\columnwidth]{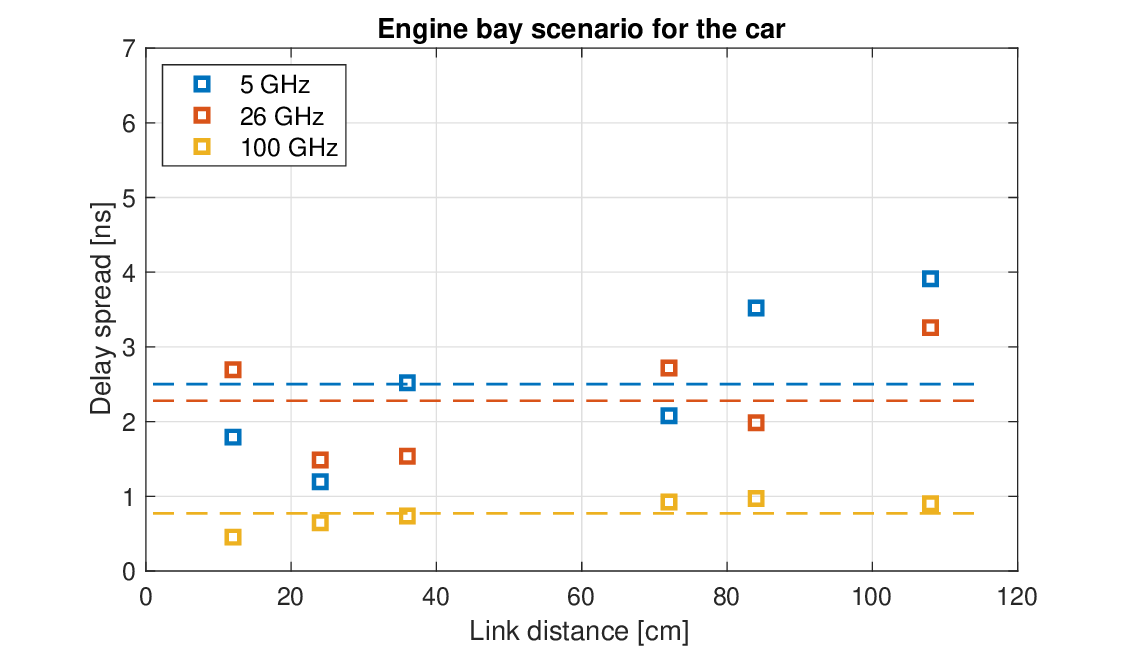}}

\caption{\added{RMS delay spread results for measurements in (a) the van and (b) car, respectively. Mean values are plotted with dashed lines.}}
\label{fig:tau}
\end{figure}

Fig. \ref{fig:tau} (a) illustrates RMS delay spread in the van ranging from \added{0.39 to 2.55 ns} for low frequency bands and from \added{0.22 to 3.22 ns} for mmWave bands. In contrast, Sub-Thz bands exhibit a smaller range of \added{0.55 to 1.30 ns}.
The average value of RMS delay spread is almost the same for low frequency and mmWave bands. The high similarity of average RMS delay spread between these two bands can be attributed to the fact that the variation trend of their CIRs are similar for the measurements in the van.
The RMS delay spread results obtained from the measurements in the car are shown in Fig. \ref{fig:tau} (b). The RMS delay spread range in this case is from \added{1.49 to 3.26 ns} with an average value of \added{2.28 ns} for low frequency bands. The range is from \added{1.20 to 3.91 ns} with an average value of \added{2.50 ns} for mmWave bands. The RMS delay spread results for Sub-THz are similar in the van and car. The pathloss significantly increases as the frequency goes to Sub-THz bands. Therefore, MPCs with large delays could have little influence on the delay spread at 100 GHz due to the low path power.

\section{Channel Spatial profiles}\label{sec:results_padp}
\begin{figure}
\centering
\subfigure []
{\includegraphics[width=1\columnwidth]{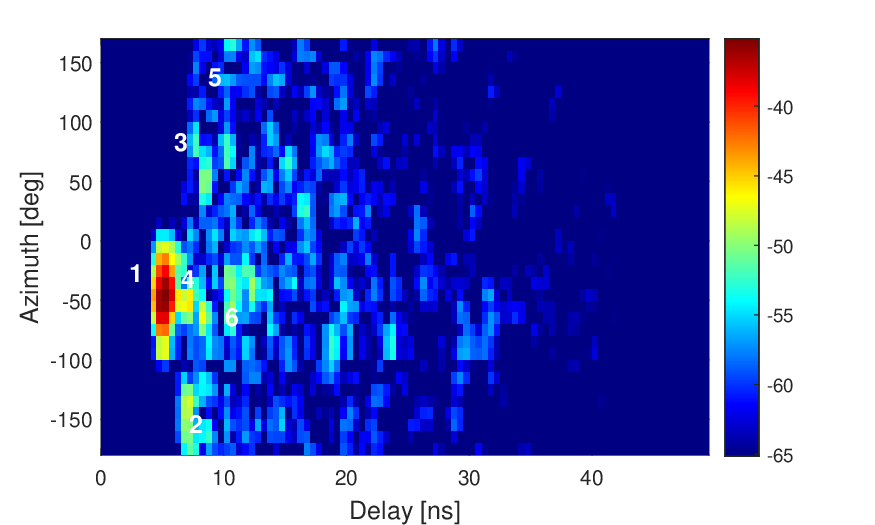}}
\subfigure []
{\includegraphics[width=1\columnwidth]{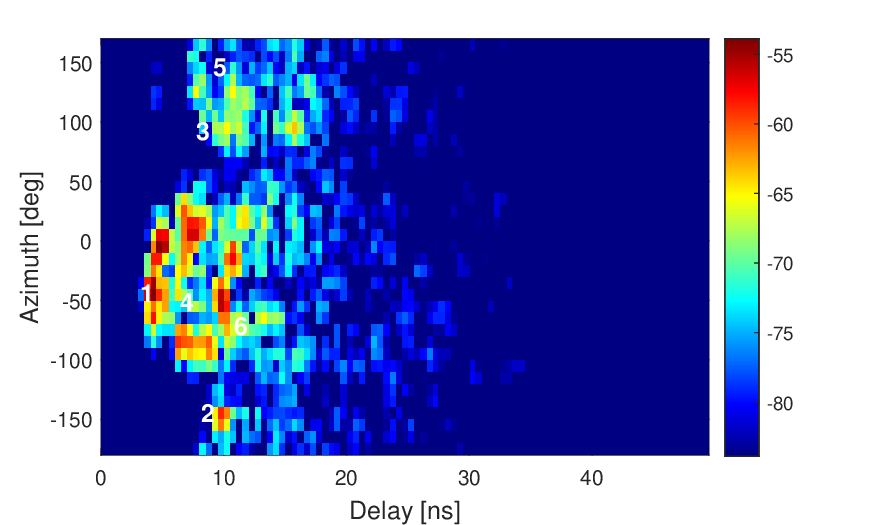}}
\subfigure []
{\includegraphics[width=1\columnwidth]{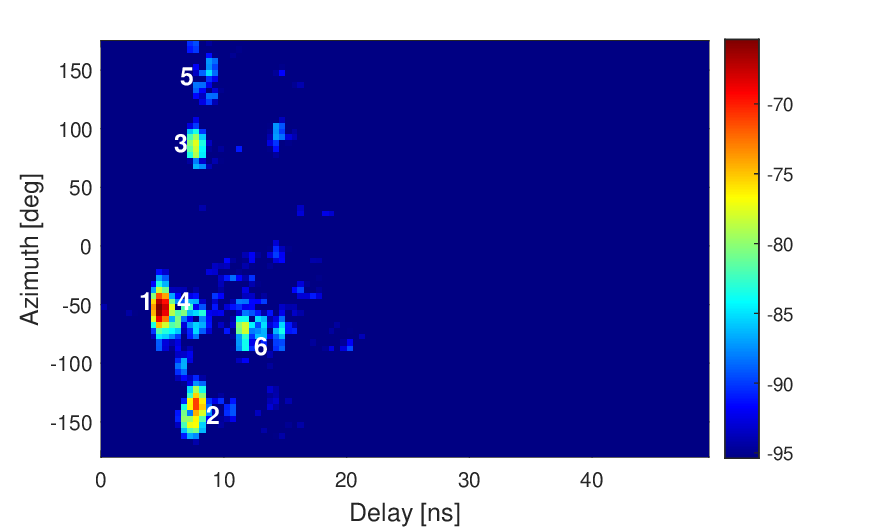}}
\caption{Measured PADPs in Scenario 1 at (a) low frequency, (b) mmWave and (c) Sub-THz bands, respectively.}
\label{fig:padp_s1}
\end{figure}

\begin{figure}
\centering
{\includegraphics[width=1\columnwidth]{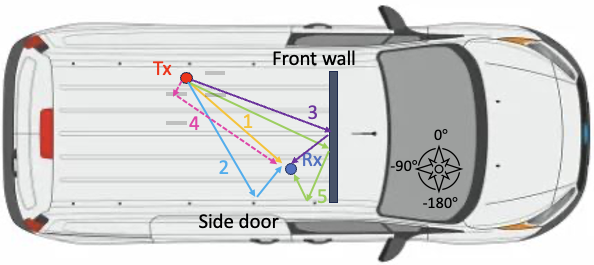}}
{\includegraphics[width=1\columnwidth]{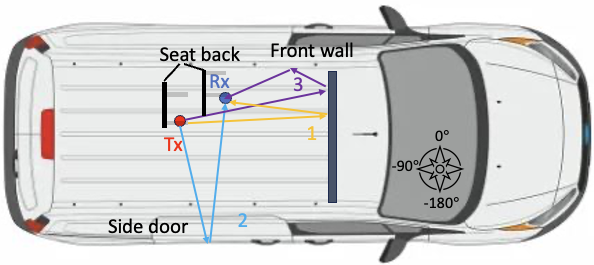}}
\caption{\added{Trajectories of the main MPCs in relation to geometry of (a) Scenario 1 and (b) Scenario 2, respectively.}}
\label{fig:trajectory_s1}
\end{figure}

\begin{figure}
\centering
\subfigure []
{\includegraphics[width=1\columnwidth]{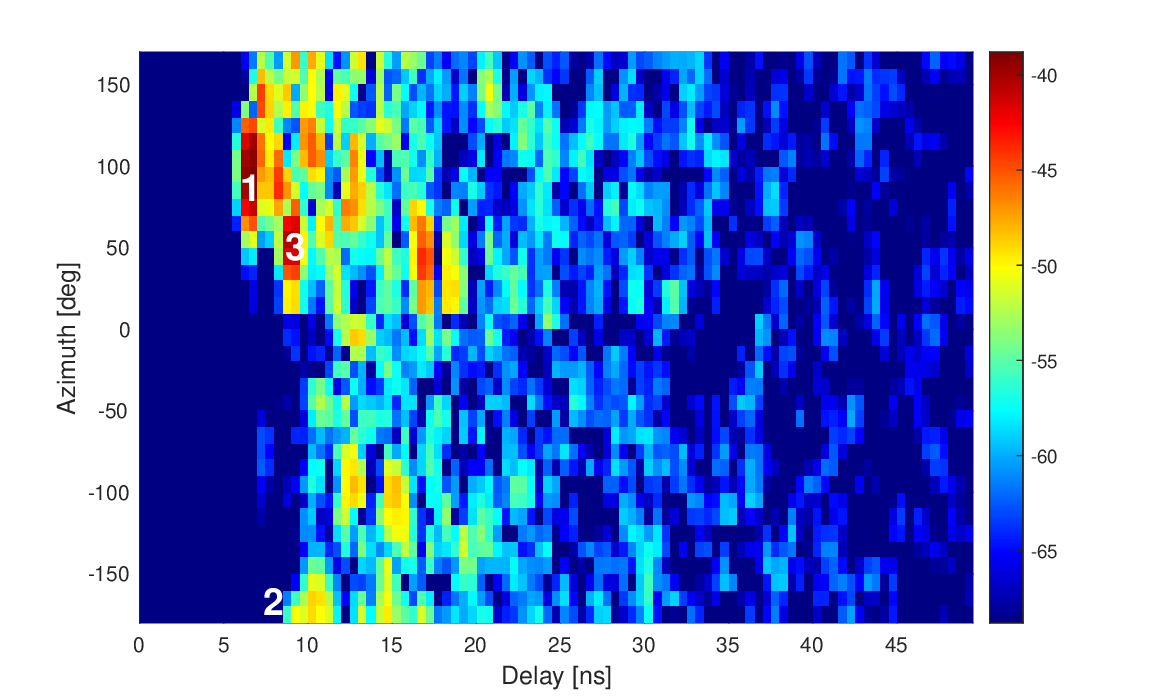}}
\subfigure []
{\includegraphics[width=1\columnwidth]{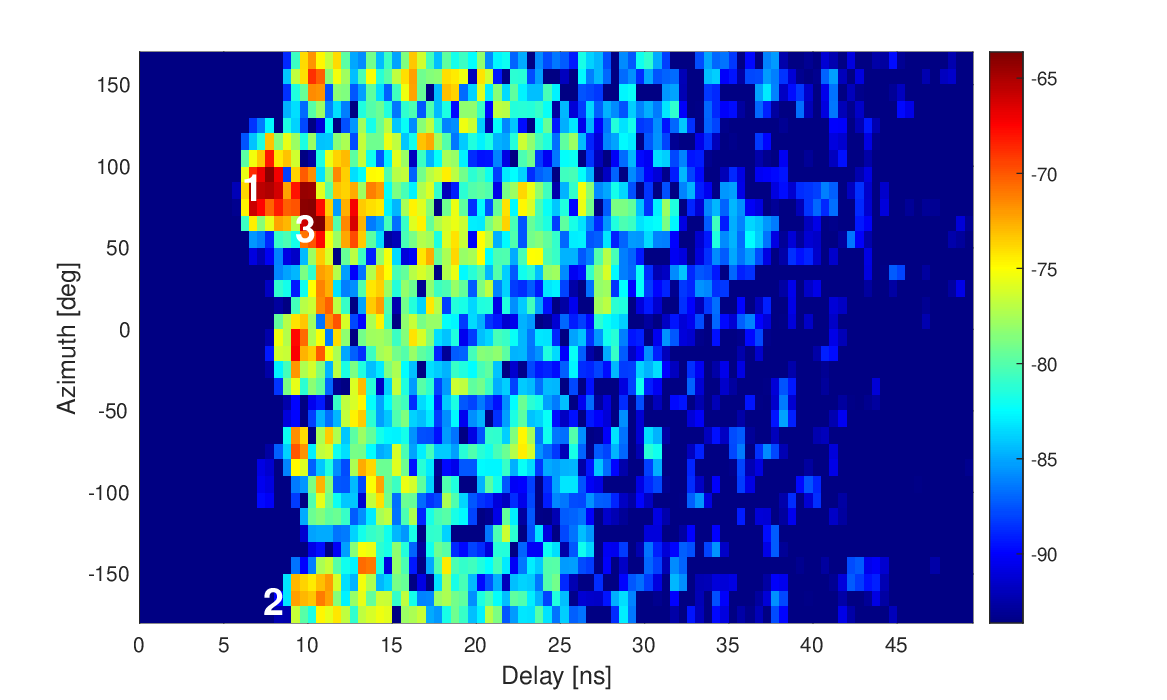}}
\caption{Measured PADPs in Scenario 2 (NLOS) at (a) low frequency, and (b) mmWave bands, respectively.}
\label{fig:padp_s2}
\end{figure}

\begin{figure}
\centering
\subfigure []
{\includegraphics[width=1\columnwidth]{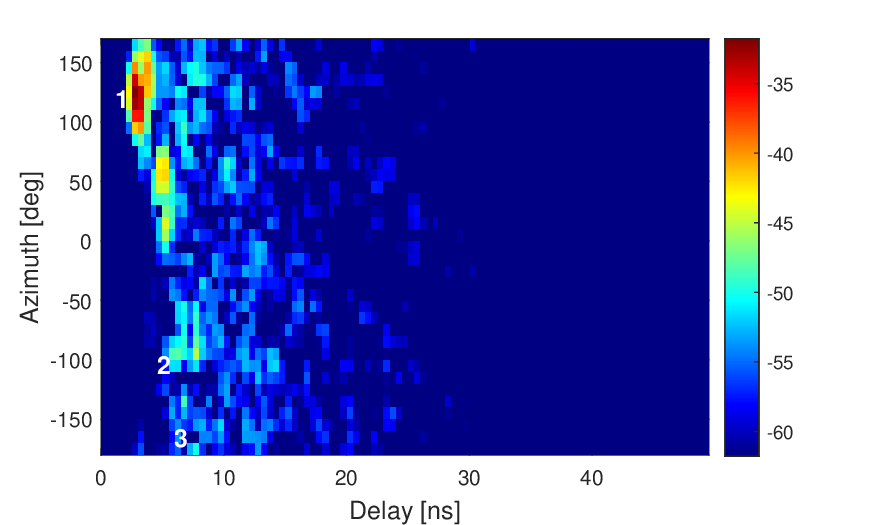}}
\subfigure []
{\includegraphics[width=1\columnwidth]{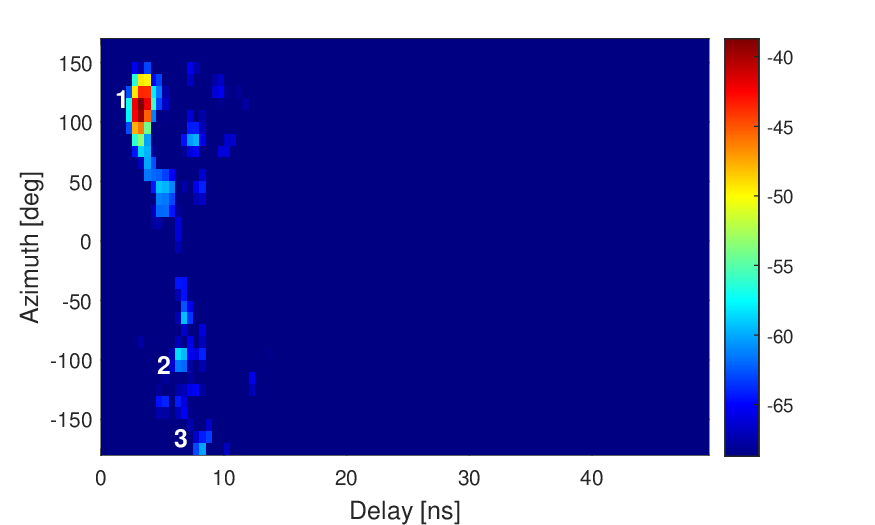}}
\subfigure []
{\includegraphics[width=1\columnwidth]{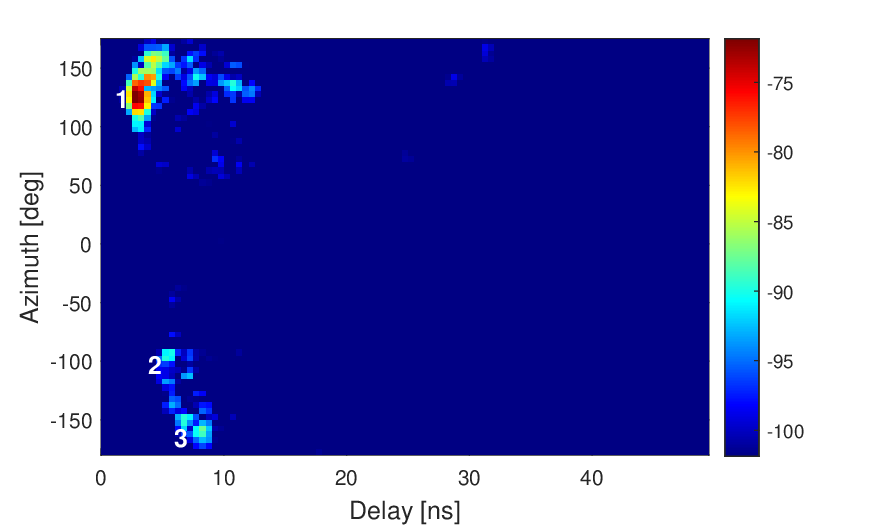}}
\caption{Measured PADPs in Scenario 3 at (a) low frequency, (b) mmWave and (c) Sub-THz bands, respectively.}
\label{fig:padp_s3}
\end{figure}

\begin{figure}
\centering
{\includegraphics[width=1\columnwidth]{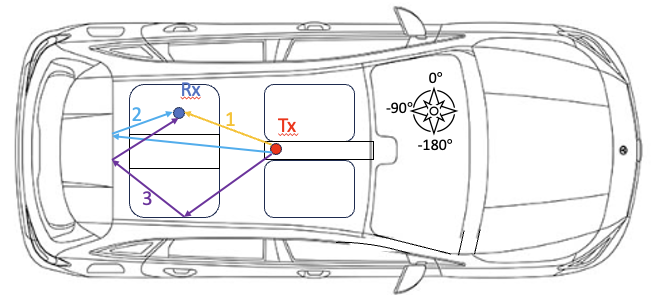}}

\caption{Trajectory of the main MPCs in relation to geometry of Scenario 3.}
\label{fig:trajectory_s2}
\end{figure}
In this section, we focus on investigating the channel spatial profiles in the passenger cabin of the vehicles. Fig. \ref{fig:padp_s1} shows the PADPs for three frequency bands measured in the Scenario 1 (as illustrated in Fig. \ref{fig:scenarios_spatial} (a)). A dynamic range of 30 dB is used in these PADPs and the antenna gains are not calibrated from the results. 
The directional antenna used as the Rx antenna at the low-frequency band has approximately twice larger HPBW compared to the other two bands, resulting in a lower angular resolution. Richer multipath environments can be observed in the low frequency bands, and more sparse channels with a few MPCs can be observed in Sub-THz bands. MPCs detected above the power threshold are distributed within a delay range of 40 ns for the previous two bands, while within a range of around 20 ns for the Sub-THz band. Specular reflection paths are surrounded by several scattering paths in low and mmWave frequency bands. Several dominant MPCs which can be identified in all the frequency bands are marked in Fig. \ref{fig:padp_s1}. 

The LOS path has an angle of arrival (AOA) in the azimuth plane of -40$^\circ$ to -50$^\circ$. 
Differences in the obtained AOA for various frequency bands arise from challenges in accurately aligning the positions of Tx and Rx antennas. The trajectory of the identified MPCs, as depicted in Fig. \ref{fig:trajectory_s1}, is derived from the geometric information and the estimated MPC parameters. Paths 2 and 3 are the single bounce reflections from the side passenger doors and front wall of the passenger cabin, respectively. Path 4 is produced by scattering from the edges of the metal chair legs, resulting in notable variations in power levels relative to LOS path power in different frequency bands. Path 5 is a double bounce reflection path from the front wall and side door of the passenger cabin. 
The parameters of the marjor MPCs in Fig. \ref{fig:padp_s1} (b) do not align perfectly with the results obtained for the low frequency and Sub-THz bands.
Stronger and more reflection paths with AOA between -100$^\circ$ to 50$^\circ$ are observed for mmWave bands.
 This inconsistency may be attributed to differences in the height of the Rx antenna used for different bands. The Rx antenna utilized for the mmWave band is positioned lower compared to the other two bands due to mounting considerations. A lower positioning of the \added{Rx} antenna results in more reflections from the metal legs of the seats and front door in the passenger cabin. 
 
 The measured PADPs of the Scenario 2 for the low frequency and the mmWave bands are shown in Fig. \ref{fig:padp_s2} (a) and (b), respectively. \added{Note that we do not provide the PADP at the Sub-THz band for NLOS scenario due to the reduced accuracy in identifying MPCs with a limited dynamic range. Several main MPCs are marked in Fig. \ref{fig:padp_s2} and the corresponding trajectory is shown in Fig. \ref{fig:trajectory_s1} (b). Strong reflections such as paths 1 and 2 are one bounce and double bounce reflections from the front wall in the passenger cabin. The path 3 is produced by the reflection from the side door.} The AOA of major MPCs for the NLOS scenario are distributed winthin the range of [0$^\circ$, 150$^\circ$]. They are mainly produced by the single and multiple bounces from the front wall of the passenger cabin. 

The PADPs measured in the passenger cabin of the car, i.e., Scenario 3 are shown in Fig. \ref{fig:padp_s3}. As can be observed in Fig.\ref{fig:scenarios_spatial} (c), the space in the passenger cabin of the car is quite limited compared to the space in the van. The Tx antenna is positioned at the center between the driver seats, while the Rx antenna is situated on the left passenger seat. 
The main MPCs are observed within a delay range of 30 ns for the low frequency band, while a range of approximately 15 ns is notable for the other two bands. Fewer MPCs are observed in Scenario 3 compared with the results obtained from Scenario 1. Three MPCs which can be detected in all the frequency bands are marked in Fig. \ref{fig:padp_s3}. The corresponding trajectory inside the car is shown in Fig. \ref{fig:trajectory_s2}. A strong LOS path followed by several weak scattering paths can be observed. Path 2 is a single bounce path reflected by the passenger chair. Path 3 is a double bounce reflection path from the side passenger door and the back seat. 
The angular spreads calculated for Scenario 1, 2, and 3 across three frequency bands are summarized in Table \ref{table:AS}. In the LOS scenario within the van, the angular spread is smaller compared to the NLOS scenario. Additionally, in Scenario 3, the angular spread decreases as the frequency increases.

\begin{table}
 \begin{center}
\caption{Summary of the Angular Spread Results}
\label{table:AS}
\setlength{\tabcolsep}{1.5pt}
\renewcommand\arraystretch{2}
\begin{tabular}{|c|c|c|c|}
\hline
Scenario No. & 4.4-6.4 GHz & 28-30 GHz  & 99-101 GHz \\ \hline
Scenario 1 & $47.3^\circ$ & $38.4^\circ$ & $42.5^\circ$ \\ \hline
Scenario 2 & $56.6^\circ$ & $70.5^\circ$ & N/A \\ \hline
Scenario 3 & $62.2^\circ$ & $56.2^\circ$ & $37.0^\circ$ \\ \hline
\end{tabular}
\end{center}
\end{table}

\section{Conclusion}\label{sec:conclusion}
\added{In-vehicle subnetworks are promising for replacing the conventional cable-based communication protocols, providing sub-millisecond latency for highly specialized use cases. In this paper, a comprehensive measurement study of channel characteristics in the engine bay and passenger cabins of a van and a car at multiple bands have been reported. The CIRs, delay spread and pathloss are analyzed for the engine bay scenario with both the FI and CI pathloss models used to fit the measured results. The results reveal that engine bay scenarios are dominated by LOS path and complex multipath propagation with major MPCs from the metal covers and adjacent engine components. Typically, fewer detectable MPCs are observed at sub-THz bands compared to the other bands. The RMS delay spread results at sub-THz frequencies are smaller than those at the other two bands, and the values across all three bands are lower than those reported in an indoor office and factory environment due to the short range transmission. The calculated pathloss exponents across all three frequency bands, for both the FI and CI models, are all smaller than 2, likely due to the tight propagation space in the engine bay. The PADP and angular spread results are extracted and discussed for passenger cabin scenarios. The spatial-temporal characteristics of the strong paths across the three bands generally align well with each other. As anticipated, sparse channels are observed at sub-THz frequency bands.}

\added{Based on the above findings, the signals at around 5 GHz and 26 GHz bands, both with strong LOS path and MPCs, are likely to support transmission within short distance whereas Sub-THz signals encounter significant attenuation and may need multiple access point antennas distributed across the area or a larger antenna array to compensate for the high losses. Millimeter wave bands around 26 GHz show superior properties for fulfilling demanding use cases by offering abundant spectrum resources and a sufficient strong link budget at the same time. In addition, understanding the channel characteristics of interference from nearby subnetworks and external wide area cells is critical for maintaining high reliability of in-vehicle subnetworks. Interference channel models for the prospective in-vehicle subnetworks also need to be explored, which might be addressed in our future research.}

\section*{Acknowledgments}

This research is supported by HORIZON-JU-SNS-2022-STREAM-B-01-03 6G-SHINE project (Grant Agreement No. 101095738).

\bibliographystyle{IEEEtran}
\addcontentsline{toc}{section}{\refname}\bibliography{main}

\end{document}